\documentclass[a4paper,11pt,DIV=12]{scrartcl}
\pdfoutput=1

\usepackage[ttscale=0.9]{libertine}
\usepackage{mathrsfs}
\usepackage[intlimits]{amsmath}
\usepackage{amssymb}
\usepackage{slashed}
\usepackage[titletoc,title]{appendix}
\usepackage[affil-it]{authblk}
\usepackage[numbers,sort&compress]{natbib}
\usepackage{graphicx}
\usepackage[normalem]{ulem}
\usepackage{layouts}
\usepackage{bbold}
\usepackage[protrusion=true,expansion,kerning=true,tracking=true,final]{microtype}
\usepackage[
            colorlinks=true,
            linkcolor=hblue,
            citecolor=hgreen,
            filecolor=hblue,
            urlcolor=hred
            ]{hyperref}
\usepackage[dvipsnames,x11names]{xcolor}
\definecolor{hgreen}{rgb}{0,.3,0}
\definecolor{hred}{rgb}{.3,0,0}
\definecolor{orange}{rgb}{1,0.5,0}
\definecolor{hblue}{rgb}{0,0,.3}
\definecolor{LightGray}{gray}{0.95}
\definecolor{gray}{gray}{0.6}

\usepackage[T1,LY1]{fontenc}

\makeatletter
\DeclareOldFontCommand{\rm}{\normalfont\rmfamily}{\mathrm}
\DeclareOldFontCommand{\sf}{\normalfont\sffamily}{\mathsf}
\DeclareOldFontCommand{\tt}{\normalfont\ttfamily}{\mathtt}
\DeclareOldFontCommand{\bf}{\normalfont\bfseries}{\mathbf}
\DeclareOldFontCommand{\it}{\normalfont\itshape}{\mathit}
\DeclareOldFontCommand{\sl}{\normalfont\slshape}{\@nomath\sl}
\DeclareOldFontCommand{\sc}{\normalfont\scshape}{\@nomath\sc}

\usepackage{scrlayer-scrpage}
\allowdisplaybreaks
\setcapindent{1em}

\setkomafont{captionlabel}{\bfseries}
\setkomafont{caption}{\itshape}
\setkomafont{titlehead}{\normalsize\sffamily}
\addtokomafont{title}{\boldmath}
\addtokomafont{section}{\boldmath}

\KOMAoptions{headinclude=false,
             footinclude=false,
             twoside=false,
             parskip=false,
             draft=false,
             abstract=true,
             numbers=noenddot,
             DIV=12}

\definecolor{Blu}{rgb}{0.,0.,1.}
\definecolor{Red}{rgb}{1.,0.,0.}
\definecolor{Darkgreen}{rgb}{0,0.5,0.}
\definecolor{Purple}{rgb}{0.5,0.,0.5}

\begin{document}

\title{Four-loop QCD mixing of current-current operators}
\date{\today}
\renewcommand\Authands{, }
\author[a]{Joachim~Brod%
        \thanks{\texttt{joachim.brod@uc.edu}}}
\author[b]{Emmanuel~Stamou%
        \thanks{\texttt{emmanuel.stamou@tu-dortmund.de}}}
\author[b]{Tom~Steudtner%
        \thanks{\texttt{tom2.steudtner@tu-dortmund.de}}}

\affil[a]{{\large Department of Physics, University of Cincinnati, Cincinnati, OH 45221, USA}}
\affil[b]{{\large Department of Physics, TU Dortmund, D-44221 Dortmund, Germany}}

\maketitle

\begin{abstract}
  We calculate the anomalous dimension of the $|\Delta S| = 1$
  current-current operators of the weak effective Lagrangian at
  next-to-next-to-next-to-leading order (NNNLO) in QCD. This
  constitutes the first step towards a full four-loop calculation of
  the QCD correction to $\epsilon_K$, the measure for indirect CP
  violation in the neutral kaon system. We present fully analytic
  results, together
  with the expressions necessary to transform our results to a basis
  with an arbitrary different definition of evanescent operators. As
  an application, we calculate the corresponding results in the
  ``standard'' operator basis used in $B$-physics applications.
\end{abstract}

\setcounter{page}{1}

\section{Introduction\label{sec:introduction}}

Indirect CP violation in the neutral kaon system has been measured
with extraordinary precision: the current experimental value,
$|\epsilon_K| = 2.228(11) \times 10^{-3}$, has an uncertainty of about
four permil~\cite{ParticleDataGroup:2024cfk}. This measurement can be
viewed as a precision test of the standard model (SM). By contrast,
the SM prediction currently has a (non-parametric) theory uncertainty
of the order of a few percent~\cite{Brod:2021hsj}. To fully exploit
the sensitivity of $\epsilon_K$ to new dynamics above the weak scale,
the theory uncertainty must be reduced by an order of magnitude. The
non-perturbative contribution to the SM prediction, namely, the
hadronic matrix element $B_K$ of the local $|\Delta S| = 2$ operator,
can be computed using lattice QCD methods. Current results have an
uncertainty of about one percent~\cite{FlavourLatticeAveragingGroupFLAG:2024oxs,
Boyle:2024gge}, and are expected to improve further in the near
future, not least because of the inclusion of higher-order terms in
the conversion between the perturbative and non-perturbative
renormalization schemes~\cite{Gorbahn:2024qpe}.

However, presently the largest room for improvement lies in the
perturbative contributions to $\epsilon_K$. It is convenient to split
the perturbative part of the SM prediction into the top-quark
contribution and the charm-top contribution~\cite{Christ:2012se,Brod:2019rzc}.
The top-quark contribution has been calculated at
next-to-leading order (NLO) in QCD~\cite{Buras:1990fn}, and at NLO in
the electroweak interaction~\cite{Brod:2021qvc}, 
with a remaining
uncertainty at the percent level. The charm-top contribution is known
up to next-to-next-to-leading order (NNLO) in
QCD~\cite{Herrlich:1993yv, Herrlich:1996vf, Brod:2010mj, Brod:2019rzc}
and NLO in the electroweak interaction~\cite{Brod:2022har}, again with
a residual uncertainty at the percent level. Part of the
next-to-next-to-next-to-leading-order (NNNLO) QCD corrections to the
charm-top contribution, namely, the three-loop matching conditions at
the charm-quark scale, are also known~\cite{Brod:2011ty}.

In this work, we take the next step towards the full four-loop
renormalization-group (RG) analysis of the charm-top contributions to
$\epsilon_K$. The four-loop mixing in the $|\Delta S| = 2$ sector,
as well as all relevant three-loop matching conditions, will be
presented in a forthcoming publication. Here, we present as a
preliminary step the four-loop anomalous dimension of the $|\Delta S|
= 1$ current-current operators. The corresponding three-loop anomalous
dimensions and the corresponding two-loop initial conditions have been
calculated nearly thirty years ago~\cite{Chetyrkin:1997gb,
  Gorbahn:2004my, Bobeth:1999mk}.

Our computational setup is based on the \texttt{MaRTIn}
code~\cite{Brod:2024zaz}, written in
\texttt{FORM}~\cite{Vermaseren:2000nd, Davies:2026cci}. This setup has
been extended for the calculation of four-loop vacuum diagrams using
the code \texttt{FMFT}~\cite{Pikelner:2017tgv}, and master integrals
calculated in Ref.~\cite{Czakon:2004bu}. The diagrams have been
generated using \texttt{qgraf}~\cite{Nogueira:1991ex}. We employ the
infrared rearrangement algorithm presented in
Ref.~\cite{Chetyrkin:1997fm} to extract the ultraviolet divergences.

This article is organized as follows. 
In Sec.~\ref{sec:results}, we present the operator basis, as well as 
the results of our calculation and discuss various checks on our calculation. 
In Sec.~\ref{sec:scheme} we show how to convert our result to an operator
basis using different definitions of evanescent operators, and give
the results in the basis of Ref.~\cite{Gorbahn:2004my}. We conclude in
Sec.~\ref{sec:conclusions}. In App.~\ref{sec:Z} we present the results
for various renormalization constants that are needed in intermediate
steps of the calculation and to perform a change of basis.

\section{Operator basis and anomalous dimension\label{sec:results}}

We consider the effective Lagrangian relevant for $|\Delta S| = 1$
transitions in the SM,
\begin{equation}\label{eq:lag:s1}
\begin{split}
  \mathcal{L}^{|\Delta S|=1}
  = - \frac{4 G_{\rm F}}{\sqrt{2}}
  \sum_{q,q'=u,c} \!\!\! V_{qs}^\ast V_{q'd} (C_{+} Q_+^{qq'} + C_{-} Q_-^{qq'}) + \text{h.c.} \,,
\end{split}
\end{equation}
with $G_{\rm F}$ the Fermi constant and $V_{ij}$ the Cabibbo-Kobayashi-Maskawa-matrix elements. 
The operators $Q_\pm^{qq'}$
are defined as the linear combinations of
\begin{align}
  Q_1^{qq'} & = (\bar s_L \gamma_{\mu} T^a q_L)\otimes 
        (\bar q_L' \gamma^{\mu} T^a d_L) \,, \\
  Q_2^{qq'} & = (\bar s_L \gamma_{\mu} q_L)\otimes
        (\bar q_L' \gamma^{\mu} d_L) \,,
\end{align}
that do not mix under the QCD RG evolution at leading order, namely,
\begin{equation}
  Q_\pm^{qq'}
  = \frac{1}{2} \bigg( 1 \pm \frac{1}{N}\bigg) Q_2^{qq'} \pm Q_1^{qq'} \,,
\end{equation}
with $N=3$ the number of colors.

In the context of dimensional regularization, the Lagrangian in
Eq.~\eqref{eq:lag:s1} must be supplemented by evanescent
operators. They arise in intermediate stages of calculations because
certain relations (such as Dirac algebra and Fierz transformations)
are only valid in four space-time dimensions.  Thus, evanescent
operators are required to absorb all UV divergences.  We define them
in such a way that the anomalous dimension in the physical sector
remains diagonal up to NNNLO for $N_f$ active quark flavors
\begin{align}
  E_1^{qq'(1)} & = (\bar s_L \gamma_{\mu_1\mu_2\mu_3} T^a q_L)\otimes 
        (\bar q_L' \gamma^{\mu_1\mu_2\mu_3} T^a d_L)
         - \big(16 - 4\epsilon - 4\epsilon^2 \big) Q_1^{qq'} \,,\\
  E_2^{qq'(1)} & = (\bar s_L \gamma_{\mu_1\mu_2\mu_3} q_L)\otimes
        (\bar q_L' \gamma^{\mu_1\mu_2\mu_3} d_L)
         - \big(16 - 4\epsilon - 4\epsilon^2 \big) Q_2^{qq'} \,,\\
  E_1^{qq'(2)} & = (\bar s_L \gamma_{\mu_1\mu_2\mu_3\mu_4\mu_5} T^a q_L)\otimes
        (\bar q_L' \gamma^{\mu_1\mu_2\mu_3\mu_4\mu_5} T^a d_L)
                 - \bigg(256 - 224\epsilon - \frac{5712}{25}\epsilon^2 \bigg) Q_1^{qq'} \,, \\
  E_2^{qq'(2)} & = (\bar s_L \gamma_{\mu_1\mu_2\mu_3\mu_4\mu_5} q_L)\otimes
        (\bar q_L' \gamma^{\mu_1\mu_2\mu_3\mu_4\mu_5} d_L)
         - \bigg(256 - 224\epsilon - \frac{10032}{25}\epsilon^2 \bigg) Q_2^{qq'} \,, \\
\begin{split}
  E_1^{qq'(3)} & = (\bar s_L \gamma_{\mu_1\mu_2\mu_3\mu_4\mu_5\mu_6\mu_7} T^a q_L)\otimes
        (\bar q_L' \gamma^{\mu_1\mu_2\mu_3\mu_4\mu_5\mu_6\mu_7} T^a d_L) \\
& \quad
         - \bigg[4096 - \bigg(\frac{404096}{45} + \frac{5888}{15} N_f\bigg) \epsilon
                 + \bigg(\frac{2692488832}{8725} + \frac{12305408}{1047} N_f\bigg) \epsilon^2 \bigg] Q_1^{qq'} \,,
\end{split}\\
\begin{split}
  E_2^{qq'(3)} & = (\bar s_L \gamma_{\mu_1\mu_2\mu_3\mu_4\mu_5\mu_6\mu_7} q_L)\otimes
        (\bar q_L' \gamma^{\mu_1\mu_2\mu_3\mu_4\mu_5\mu_6\mu_7} d_L) \\
& \quad
         - \bigg[4096 - \bigg(\frac{42112}{9} - \frac{4352}{3} N_f\bigg) \epsilon
                 + \bigg( \frac{5106291712}{8725} + \frac{132105472}{5235} N_f\bigg) \epsilon^2 \bigg] Q_2^{qq'} \,,
\end{split}\\
  E_1^{qq'(4)} & = (\bar s_L \gamma_{\mu_1\mu_2\mu_3\mu_4\mu_5\mu_6\mu_7\mu_8\mu_9} T^a q_L)\otimes
        (\bar q_L' \gamma^{\mu_1\mu_2\mu_3\mu_4\mu_5\mu_6\mu_7\mu_8\mu_9} T^a d_L)
         - 65536 Q_1^{qq'} \,, \\
  E_2^{qq'(4)} & = (\bar s_L \gamma_{\mu_1\mu_2\mu_3\mu_4\mu_5\mu_6\mu_7\mu_8\mu_9} q_L)\otimes
        (\bar q_L' \gamma^{\mu_1\mu_2\mu_3\mu_4\mu_5\mu_6\mu_7\mu_8\mu_9} d_L)
        - 65536 Q_2^{qq'} \,,
\end{align}
where the superscript in parentheses indicates the loop-level at which
the evanescent operator first appears. We have used a shorthand
notation for products of $\gamma$ matrices:
\begin{equation}
  \gamma_{\mu_1\dots\mu_n} \equiv \gamma_{\mu_1}\dots\gamma_{\mu_n} \,.
\end{equation}
This choice of evanescent operators is not unique. With our
convention, the one- and two-loop evanescent operators are exactly the
ones defined in Ref.~\cite{Brod:2010mj}, while in the three-loop
evanescent operators we have chosen to set the coefficients of
$\epsilon^3$ to zero. The four-loop evanescent operators are only
needed for the projection of our results to the operator basis; we
could have added any power of $\epsilon$ times a linear combination of
physical operators without affecting our final results.

The RG flow of the Wilson coefficients is governed by the RG equation
\begin{equation}
  \frac{d {\vec{C}^t}}{d\log\mu} =  \vec{C}^t \gamma
\quad\text{with}\quad
\gamma =   \frac{\alpha_s}{4\pi}\gamma^{(0)}
         + \left(\frac{\alpha_s}{4\pi}\right)^2\gamma^{(1)}
         + \left(\frac{\alpha_s}{4\pi}\right)^3\gamma^{(2)}
         + \left(\frac{\alpha_s}{4\pi}\right)^4\gamma^{(3)}
         + \ldots\,,
\end{equation}
where the superscript $t$ denotes transposition and $\gamma^{(0)}$,
$\gamma^{(1)}$, $\gamma^{(2)}$, $\gamma^{(3)}$ are the one-, two-,
three-, and four-loop anomalous-dimension matrices, respectively.  The
anomalous dimension matrix in the physical sector is given in terms of
the renormalization constants by
\begin{align}
  \gamma^{(0)} & = 2 Z^{(1,1)} \,, \label{eq:adm:0} \\
  \gamma^{(1)} & = 4 Z^{(2,1)} - 2 Z^{(1,1)} Z^{(1,0)} \,, \\
  \gamma^{(2)} & = 6 Z^{(3,1)} - 4 Z^{(2,1)} Z^{(1,0)} - 2 Z^{(1,1)} Z^{(2,0)} \,, \\
  \gamma^{(3)} & = 8 Z^{(4,1)} - 6 Z^{(3,1)} Z^{(1,0)}
                  - 4 Z^{(2,1)} Z^{(2,0)}- 2 Z^{(1,1)} Z^{(3,0)} \,. \label{eq:adm:3}
\end{align}
The expressions in Eqs.~\eqref{eq:adm:0}--\eqref{eq:adm:3} are valid
if one restricts the anomalous dimension matrices to the physical
sector; otherwise, additional terms would enter. Here, the $Z$ factors
correspond to matrices that contain the renormalization constants for
the Wilson coefficients, such that $C_\text{bare} = C Z$, where $C$
are the renormalized Wilson coefficients. We write them as
coefficients in a double expansion in the strong coupling constant
$\alpha_s$ and the parameter $\epsilon$, appearing in the context of
dimensional regularization via the space-time dimension $d = 4 -
2\epsilon$:
\begin{equation}\label{eq:def:Z:exp}
  Z = \sum_k \sum_{l=0}^k \frac{\alpha_s^{k}}{(4\pi)^{k} \epsilon^l}
  Z^{(k,l)} \,.
\end{equation}
Note that the finite parts of the $Z$ factors (corresponding to $l=0$)
are only non-zero for the mixing of evanescent into physical
operators~\cite{Dugan:1990df, Herrlich:1994kh}.

By explicit calculation, we reproduce the known results in the
physical sector up to NNLO,
\begin{align}
  \gamma_{Q_+^{qq'} \to Q_+^{qq'}}^{(0)} & = 4 \,, \hspace{3cm}
  \gamma_{Q_-^{qq'} \to Q_-^{qq'}}^{(0)} = -8 \,, \\
  \gamma_{Q_+^{qq'} \to Q_+^{qq'}}^{(1)}
  & = -7+\frac{4}{9}N_f \,,                          \hspace{1.5cm}
  \gamma_{Q_-^{qq'} \to Q_-^{qq'}}^{(1)}
  = -14-\frac{8}{9}N_f \,, \\
  \gamma_{Q_+^{qq'} \to Q_+^{qq'}}^{(2)}
& = \frac{275267}{150} - \frac{52891}{675} N_f - \frac{260}{81} N_f^2
    - \bigg( 672 + \frac{160}{3} N_f \bigg) \zeta_3 \,, \\
  \gamma_{Q_-^{qq'} \to Q_-^{qq'}}^{(2)}
& = \frac{12297}{25} - \frac{62686}{675} N_f + \frac{520}{81} N_f^2
    + \bigg( 672 - \frac{320}{3} N_f \bigg) \zeta_3 \,,
\end{align}
in agreement with the literature~\cite{Brod:2010mj}; see
Refs.~\cite{Buras:1992tc, Ciuchini:1993vr, Chetyrkin:1997gb,
  Gorbahn:2004my} for the original calculations, using different bases
of evanescent operators. The three-loop result has been checked here by
explicit calculation for the first time. 
At four-loop, we find the mixing at NNNLO 
\begin{align}
\begin{split}
  \gamma_{Q_+^{qq'} \to Q_+^{qq'}}^{(3)}
& =   \frac{250492828261063}{101768400}
    - \frac{1023760733731}{19081575} N_f
    - \frac{37339436464}{6360525} N_f^2
    - \frac{68}{27} N_f^3 \\
& \quad
    - \bigg(
        \frac{1428841}{243}
      + \frac{3671912}{729} N_f
      - \frac{57952}{243} N_f^2
      - \frac{64}{27} N_f^3
      \bigg) \zeta_3 \\
& \quad
    + \bigg(
        \frac{616}{5}
      + \frac{104}{45} N_f
      - \frac{16}{27} N_f^2
      \bigg) \pi^4
    - \bigg(
        \frac{1563260}{81}
      - \frac{35920}{9} N_f
      \bigg) \zeta_5 \,,
\end{split}\\[1em]
\begin{split}
  \gamma_{Q_-^{qq'} \to Q_-^{qq'}}^{(3)}
& =   \frac{2677463242787}{50884200}
    - \frac{75148351018}{19081575} N_f
    + \frac{491735576}{6360525} N_f^2
    + \frac{136}{27} N_f^3 \\
& \quad
    + \bigg(
        \frac{2807306}{243}
      + \frac{4756048}{729} N_f
      - \frac{50624}{243} N_f^2
      - \frac{128}{27} N_f^3
      \bigg) \zeta_3 \\
& \quad
    - \bigg( 
        \frac{616}{5}
      + \frac{544}{45} N_f
      - \frac{32}{27} N_f^2
      \bigg) \pi^4
    + \bigg(
        \frac{1771960}{81}
      - \frac{53920}{9} N_f
      \bigg) \zeta_5 \,.
\end{split}
\end{align}
The off-diagonal entries of the anomalous dimension matrix vanish with
our choice of evanescent operators, i.e., $\gamma_{Q_\pm^{qq'} \to
  Q_\mp^{qq'}}^{(3)}=0$.  All results have been obtained using a fully
anticommuting $\gamma_5$ (NDR scheme), which is a consistent
prescription for this application given that no traces with $\gamma_5$
appear in our calculation.

We have performed various consistency checks on our calculation. First, we
have calculated all QCD renormalization constants up to the three-loop
level, and the quark field renormalization constant at the four-loop
level, and find agreement with results in the literature where
available~\cite{Gambino:2003zm}.
Our explicit results for the QCD
renormalization constants are listen in App.~\ref{sec:Z}. We have
performed all calculations up to the three-loop level in a generalized
$R_\xi$ gauge, and verified explicitly that the anomalous dimensions
are independent of the gauge parameter to that order. Due to the
complexity of the involved integrals, we have calculated all four-loop
results in `t~Hooft-Feynman gauge, i.e., for $\xi = 1$.

The finiteness of the anomalous dimension matrix implies the following
relations between the renormalization constants:
\begin{align}
  Z^{(2,2)} =~ & 
               \frac{1}{2} Z^{(1,1)} Z^{(1,1)} 
             - \frac{1}{2} \beta_0 Z^{(1,1)}
             \,, \label{eq:ch:z22} \\[0.5em] 
  Z^{(3,3)} =~ & 
               \frac{1}{6} Z^{(1,1)} Z^{(1,1)} Z^{(1,1)} 
             - \frac{1}{2} \beta_0 Z^{(1,1)} Z^{(1,1)}
             + \frac{1}{3} \beta_0^2 Z^{(1,1)}
             \,, \label{eq:ch:z33} \\[0.5em]
\begin{split} 
  Z^{(3,2)} =~ & 
             - \frac{1}{6} Z^{(1,0)} Z^{(1,1)} Z^{(1,1)}
             - \frac{1}{3} Z^{(1,1)} Z^{(1,0)} Z^{(1,1)}
             + \frac{2}{3}Z^{(2,1)}Z^{(1,1)} 
             \label{eq:ch:z32} \\
            & 
             + \frac{1}{3} Z^{(1,1)} Z^{(2,1)}
             + \frac{1}{6} \beta_0 Z^{(1,0)}Z^{(1,1)}
             - \frac{1}{3} \beta_1 Z^{(1,1)}
             - \frac{2}{3} \beta_0 Z^{(2,1)}
             \,,
\end{split} \\[0.5em]
\begin{split} \label{eq:ch:z44}
Z^{(4,4)} =~ &
             \frac{1}{24} Z^{(1,1)} Z^{(1,1)} Z^{(1,1)} Z^{(1,1)} 
             - \frac{1}{4} \beta_0 Z^{(1,1)} Z^{(1,1)} Z^{(1,1)}
             \\
            &
             + \frac{11}{24} \beta_0^2 Z^{(1,1)} Z^{(1,1)}
             - \frac{1}{4} \beta_0^3 Z^{(1,1)}
             \,,
\end{split} \\[0.5em]
\begin{split} \label{eq:ch:z43}
Z^{(4,3)}  =~ & 
            - \frac{1}{12} Z^{(1,0)} Z^{(1,1)} Z^{(1,1)} Z^{(1,1)}
            - \frac{1}{12} Z^{(1,1)} Z^{(1,1)} Z^{(1,0)} Z^{(1,1)}
            \\
          & 
            - \frac{1}{6} Z^{(1,1)} Z^{(1,0)} Z^{(1,1)} Z^{(1,1)}
            +  \frac{1}{4} Z^{(2,1)} Z^{(1,1)} Z^{(1,1)}
            + \frac{1}{6} Z^{(1,1)} Z^{(2,1)} Z^{(1,1)}
            \\
         & 
            + \frac{1}{12} Z^{(1,1)} Z^{(1,1)} Z^{(2,1)}
            + \frac{5}{12} \beta_0 Z^{(1,1)} Z^{(1,0)} Z^{(1,1)}
            + \frac{1}{4} \beta_0 Z^{(1,0)} Z^{(1,1)} Z^{(1,1)}
            \\
         & 
            - \frac{3}{4} \beta_0 Z^{(2,1)} Z^{(1,1)}
            - \frac{5}{12} \beta_0 Z^{(1,1)} Z^{(2,1)}
            - \frac{1}{6} \beta_0^2 Z^{(1,0)} Z^{(1,1)}
            \\
         &  
            - \frac{1}{3} \beta_1 Z^{(1,1)} Z^{(1,1)}
            + \frac{1}{2} \beta_0^2 Z^{(2,1)} 
            + \frac{1}{2} \beta_0 \beta_1 Z^{(1,1)} 
            \,,
\end{split} \\[0.5em]
\begin{split} \label{eq:ch:z42}
Z^{(4,2)}  =~ & 
            \frac{1}{4} Z^{(1,1)} Z^{(1,0)} Z^{(1,0)} Z^{(1,1)}
          + \frac{1}{6} Z^{(1,0)} Z^{(1,1)} Z^{(1,0)} Z^{(1,1)}
          + \frac{1}{12} Z^{(1,0)} Z^{(1,0)} Z^{(1,1)} Z^{(1,1)}
          \\
         & 
          - \frac{1}{2} Z^{(2,1)} Z^{(1,0)} Z^{(1,1)}
          - \frac{1}{12} Z^{(1,0)} Z^{(2,1)} Z^{(1,1)}
          - \frac{1}{6} Z^{(1,0)} Z^{(1,1)} Z^{(2,1)}
          \\
         &
          - \frac{1}{4} Z^{(1,1)} Z^{(1,0)} Z^{(2,1)}
          - \frac{1}{4} Z^{(2,0)} Z^{(1,1)} Z^{(1,1)}
          - \frac{1}{4} Z^{(1,1)} Z^{(2,0)} Z^{(1,1)}
          \\
         & 
          - \frac{1}{12} \beta_0 Z^{(1,0)} Z^{(1,0)} Z^{(1,1)}
          +  \frac{3}{4} Z^{(3,1)} Z^{(1,1)}
          + \frac{1}{4} Z^{(1,1)} Z^{(3,1)}
          + \frac{1}{2} Z^{(2,1)} Z^{(2,1)}
          \\
         &
          + \frac{1}{12} \beta_0 Z^{(1,0)} Z^{(2,1)}
          + \frac{1}{4} \beta_0 Z^{(2,0)} Z^{(1,1)}
          + \frac{1}{6} \beta_1 Z^{(1,0)} Z^{(1,1)}
          \\
         &
          - \frac{3}{4} \beta_0 Z^{(3,1)}
          - \frac{1}{2} \beta_1 Z^{(2,1)}
          - \frac{1}{4} \beta_2 Z^{(1,1)}
          \,,
\end{split}
\end{align}
where the coefficients of the QCD beta function are given by
\begin{equation}
  \beta_0 = 11 - \frac{2}{3} N_f \,, \qquad
  \beta_1 = 102 - \frac{38}{3} N_f \,, \qquad
  \beta_2 = \frac{2857}{2} - \frac{5033}{18} N_f + \frac{325}{54} N_f^2 \,.
\end{equation}
We have checked by explicit calculation that all poles up to four-loop
are in agreement with the
relations in Eqs.~\eqref{eq:ch:z22}--\eqref{eq:ch:z42}.

\section{Change of basis\label{sec:scheme}}

Having the application to kaon physics in mind, we have constructed a
basis of evanescent operators for which the anomalous-dimension matrix
for the physical operators $Q_\pm^{qq'}$ is diagonal up to NNNLO.  In
the context of $B$ physics, different but equivalent operator bases
have been used in the literature, see, e.g.,
Refs.~\cite{Chetyrkin:1996vx, Gorbahn:2004my}. In this section, we
give the expression necessary to transform our results into a basis
with different definitions of evanescent operators, and explicitly
give the results for the basis in Ref.~\cite{Gorbahn:2004my}. Our
transformation rules generalize part of the scheme transformation
relations in Ref.~\cite{Brod:2010mj} to the four-loop level.

Adopting the notation of Ref.~\cite{Brod:2010mj}, we denote the
rotation that takes the $Q_+$-$Q_-$ basis to the $Q_1$-$Q_2$ basis by
the matrix $R$, transformations of the evanescent operators among
themselves are described by the matrix $M$, and the addition of
multiples of powers of $\epsilon$ times physical operators to the
evanescent operators by the matrices $U$, $V$, and $W$,
respectively. In total we have the following transformation among all
physical ($Q$) and evanescent ($E$) operators:
\begin{equation}\label{eq:gentmat}
\begin{pmatrix}
Q'\\
E'
\end{pmatrix}
=
\begin{pmatrix}
R&0\\
0&M
\end{pmatrix}
\begin{pmatrix}
1&0\\
\epsilon U + \epsilon^2 V + \epsilon^3 W &
\end{pmatrix}
\begin{pmatrix}
Q\\
E
\end{pmatrix} \,.
\end{equation}
If we consider the renormalized Green's functions that enter the
computation of the renormalization constants $Z^{(m,n)}$ in the
original basis and apply the basis change in Eq.~\eqref{eq:gentmat},
we see that the renormalization constants of the original basis
subtract also terms that are finite and higher-order in $\epsilon$,
due to the presence of the $U$, $V$, and $W$ matrices.  This naive
rotation thus leads to an unwanted departure from the $\overline{\rm
  MS}$ scheme. Therefore, a naive rotation of the mixing constants or
the anomalous-dimension matrix does not lead to the correct
$\overline{\rm MS}$ quantities in the new basis.

To correctly obtain the $\overline{\rm MS}$ renormalization constants
in the new basis, the rotation must be accompanied by finite
renormalization constants, i.e., a change of the renormalization
scheme.  Under the rotation and the change of scheme the Wilson
coefficients transform as:
\begin{align}
\label{eq:Crotation}
\begin{pmatrix}
  C_Q  &  C_E
\end{pmatrix}
\rightarrow
\begin{pmatrix}
  C'_Q & C'_E
\end{pmatrix}
=
\begin{pmatrix}
C_Q & C_E
\end{pmatrix}
\begin{pmatrix}
  R^{-1} & 0\\
  0 & M^{-1}
\end{pmatrix}
\begin{pmatrix}
  Z'_{QQ} & Z'_{QE}\\
  Z'_{EQ} & Z'_{EE}
\end{pmatrix}\,.
\end{align}
The new renormalization constants $Z_{XY}'$ with $X=\{Q,E\}$ define
the change of scheme and admit a double expansion in terms of
$\alpha_s$ and positive powers of $\epsilon$,
\begin{align}
  Z'_{XY}  = \mathbb{1} + \sum_{k=0}^\infty\sum_{l=0}^\infty \frac{\alpha_s^k}{(4\pi)^k} \epsilon^l Z_{XY}^{\prime\,(k,-l)}\,.
\end{align}
At each order in perturbation we can determine all
$Z_{XY}^{\prime\,(k,l)}$ such that no finite or higher in $\epsilon$
terms are being subtracted by local counterterms in the rotated
renormalized Green's functions.\footnote{In accordance with the
$\overline{\rm{MS}}$ prescription, when computing the mixing of
evanescent into physical operators the finite terms of the amplitudes
(proportional to $\epsilon^0$) are subtracted by finite
renormalization constants. Thus no additional subtraction in terms of
$Z'$ constants must be imposed for these $\epsilon^0$ terms, i.e.,
$Z_{EQ}'^{(k,0)} = 0$.}  This imposes the $\overline{\rm{MS}}$
prescription for the rotated amplitude and determines all
$Z_{XY}^{\prime\,(k,l)}$ in terms of the mixing constants in the
original basis.

We find that the basis change of Eq.~\eqref{eq:gentmat} implies the finite renormalization 
of the physical operators given by
\begin{align}
  Z_{QQ}'^{(1,0)} & = - R Z_{QE}^{(1,1)} U R^{-1} \,, \label{eq:fin:z:1} \\[0.5em]
  Z_{QQ}'^{(2,0)} &= - R \left ( Z^{(2,1)}_{QE} U + Z^{(2,2)}_{QE} V -
    Z^{(1,1)}_{QE} V Z^{(1,1)}_{QQ} \right ) R^{-1} \,,\label{eq:fin:z:2} \\[0.5em]
\begin{split}
  Z_{QQ}'^{(3,0)} &= - R \bigg ( 
          Z_{QE}^{(3,1)} U
          + Z_{QE}^{(3,2)} V
          + Z_{QE}^{(3,3)} W
          - Z_{QE}^{(2,1)} V Z_{QQ}^{(1,1)}
          - Z_{QE}^{(2,2)} W Z_{QQ}^{(1,1)}
\\[0.5em]&\qquad\qquad\!\!
          + Z_{QE}^{(1,1)} V Z_{QE}^{(2,2)} U
          - Z_{QE}^{(1,1)} W Z_{QQ}^{(2,2)}
          - Z_{QE}^{(1,1)} V Z_{QQ}^{(2,1)}
\\[0.5em]&\qquad\qquad\!\!
          + Z_{QE}^{(1,1)} V Z_{QE}^{(1,1)} Z_{EQ}^{(1,0)}
          - Z_{QE}^{(1,1)} V Z_{QE}^{(1,1)} Z_{EE}^{(1,1)} U
          + Z_{QE}^{(1,1)} W Z_{QQ}^{(1,1)} Z_{QQ}^{(1,1)}
\bigg ) R^{-1} \,,
\end{split}
\label{eq:fin:z:3} 
\end{align}
where
\begin{align}
\begin{split} 
Z^{(2,2)}_{QE} & = \frac{1}{2} \left ( Z^{(1,1)}_{QE} Z^{(1,1)}_{EE} +
                                     \frac{1}{2} \gamma^{(0)} Z^{(1,1)}_{QE} - \beta_0 Z^{(1,1)}_{QE}
                             \right ) \,, 
\end{split}\\
\begin{split}                              
Z_{QE}^{(3,3)} & =   \frac{1}{24} \big(\gamma^{(0)}\big)^2 Z_{QE}^{(1,1)}
                 + \frac{1}{12} \gamma^{(0)} Z_{QE}^{(1,1)} Z_{EE}^{(1,1)}
                 + \frac{1}{6} Z_{QE}^{(1,1)} \big(Z_{EE}^{(1,1)}\big)^2 \\
& \quad
                 - \frac{1}{4} \beta_0 \gamma^{(0)} Z_{QE}^{(1,1)}
                 - \frac{1}{2} \beta_0 Z_{QE}^{(1,1)} Z_{EE}^{(1,1)}
                 + \frac{1}{3} \beta_0^2 Z_{QE}^{(1,1)} \,,
\end{split}\\
\begin{split} 
Z_{QE}^{(3,2)} & =   \frac{1}{6} \gamma^{(1)} Z_{QE}^{(1,1)}
                 + \frac{2}{3} Z_{QE}^{(2,1)} Z_{EE}^{(1,1)}
                 + \frac{1}{6} \gamma^{(0)} Z_{QE}^{(2,1)}
                 + \frac{1}{3} Z_{QE}^{(1,1)} Z_{EE}^{(2,1)} \\
& \quad
                 - \frac{1}{3} \beta_1 Z_{QE}^{(1,1)}
                 - \frac{2}{3} \beta_0 Z_{QE}^{(2,1)} \,.
\end{split}
\end{align}

The one- and two-loop expressions, in Eqs.~\eqref{eq:fin:z:1} and
\eqref{eq:fin:z:2}, respectively, are in agreement with
Refs.~\cite{Gorbahn:2004my, Buras:2006gb}, while the three-loop
expression in Eq.~\eqref{eq:fin:z:3}, $Z_{QQ}'^{(3,0)}$, is presented
here for the first time.  As discussed below, only the
$Z_{QQ}'^{(l,0)}$ entries are necessary to obtain the anomalous
dimension of the physical sector in the new basis.  However,
Eq.~\eqref{eq:Crotation} must be inverted in order to rotate the
renormalised Green's functions. As a consequence also $Z'$ entries
from the evanescent sector as well as entries with higher powers than
$\epsilon^0$ must be determined in intermediate stages to derive the
final expressions for $Z_{QQ}'^{(l,0)}$'s.  We collect all $Z$ factors
that are necessary to evaluate these expressions at the end of
Sec.~\ref{sec:Z}.

Having fixed the renormalization scheme, we can also compute the 
anomalous-dimension matrices in the new basis.
The conversion relations for these read
\begin{align}\label{eq:admtrafo}
\gamma'^{(0)} &= R\gamma^{(0)}R^{-1} \,, \\
\gamma'^{(1)} &= R\gamma^{(1)}R^{-1}
                - \left[ Z_{QQ}^{\prime(1,0)},\gamma^{\prime(0)} \right]
                - 2 \beta_0 Z_{QQ}^{\prime(1,0)} \,, \\
\begin{split}
\gamma'^{(2)} &= R\gamma^{(2)}R^{-1}
                - \left[ Z_{QQ}^{\prime(2,0)},\gamma^{\prime(0)} \right]
                - \left[ Z_{QQ}^{\prime(1,0)},\gamma^{\prime(1)} \right]
                + \left[ Z_{QQ}^{\prime(1,0)},\gamma^{\prime(0)} \right] Z_{QQ}^{\prime(1,0)} \\
&\quad          - 4 \beta_0 Z_{QQ}^{\prime(2,0)}
                - 2 \beta_1 Z_{QQ}^{\prime(1,0)}
                + 2 \beta_0 \left( Z_{QQ}^{\prime(1,0)} \right)^2 \,,
\end{split}\\
\begin{split}
\gamma'^{(3)} &= R\gamma^{(3)}R^{-1}
                - \left[ Z_{QQ}^{\prime(3,0)},\gamma^{\prime(0)} \right]
                - \left[ Z_{QQ}^{\prime(2,0)},\gamma^{\prime(1)} \right]
                - \left[ Z_{QQ}^{\prime(1,0)},\gamma^{\prime(2)} \right] \\
&\quad          + \left[ Z_{QQ}^{\prime(2,0)},\gamma^{\prime(0)} \right] Z_{QQ}^{\prime(1,0)}
                + \left[ Z_{QQ}^{\prime(1,0)},\gamma^{\prime(1)} \right] Z_{QQ}^{\prime(1,0)} \\
&\quad          + \left[ Z_{QQ}^{\prime(1,0)},\gamma^{\prime(0)} \right]
                  \bigg[ Z_{QQ}^{\prime(2,0)} - \Big(Z_{QQ}^{\prime(1,0)}\Big)^2 \bigg]
                - 6 \beta_0 Z_{QQ}^{\prime(3,0)}
                - 4 \beta_1 Z_{QQ}^{\prime(2,0)}
                - 2 \beta_2 Z_{QQ}^{\prime(1,0)} \\
&\quad          + 4 \beta_0 Z_{QQ}^{\prime(2,0)} Z_{QQ}^{\prime(1,0)}
                + 2 \beta_0 Z_{QQ}^{\prime(1,0)} Z_{QQ}^{\prime(2,0)}
                - 2 \beta_0 \left( Z_{QQ}^{\prime(1,0)} \right)^3
                + 2 \beta_1 \left( Z_{QQ}^{\prime(1,0)} \right)^2 \,.
\end{split} 
\end{align}
The two-loop and three-loop relations have appeared in
Refs.~\cite{Gorbahn:2004my, Buras:2006gb} and are reproduced here. The
four-loop expression is new, and is obtained by a straightforward
generalization of the results in Ref.~\cite{Brod:2010mj}.

\bigskip

As an application, we provide the anomalous dimension using the
``standard'' operator basis defined in Ref.~\cite{Gorbahn:2004my}. For
convenience, we list the operators explicitly:
\begin{align*}
  Q_1^{qq'} & = (\bar s_L  \gamma_{\mu} T^a q_L)\otimes 
                (\bar q_L' \gamma^{\mu} T^a d_L) \,, \\
  Q_2^{qq'} & = (\bar s_L  \gamma_{\mu} q_L)\otimes
                 (\bar q_L' \gamma^{\mu} d_L) \,,\\
  \tilde{E}_1^{qq'(1)} & = (\bar s_L  \gamma_{\mu_1\mu_2\mu_3} T^a q_L)\otimes 
                           (\bar q_L' \gamma^{\mu_1\mu_2\mu_3} T^a d_L)
                           - 16 Q_1^{qq'} \,,\\
  \tilde{E}_2^{qq'(1)} & = (\bar s_L  \gamma_{\mu_1\mu_2\mu_3} q_L)\otimes
                           (\bar q_L' \gamma^{\mu_1\mu_2\mu_3} d_L)
                           - 16 Q_2^{qq'} \,,\\
  \tilde{E}_1^{qq'(2)} & = (\bar s_L  \gamma_{\mu_1\mu_2\mu_3\mu_4\mu_5} T^a q_L)\otimes
                           (\bar q_L' \gamma^{\mu_1\mu_2\mu_3\mu_4\mu_5} T^a d_L)
                           - 256 Q_1^{qq'} -20 \tilde{E}_1^{qq'(1)} \,, \\
  \tilde{E}_2^{qq'(2)} & = (\bar s_L  \gamma_{\mu_1\mu_2\mu_3\mu_4\mu_5} q_L)\otimes
                           (\bar q_L' \gamma^{\mu_1\mu_2\mu_3\mu_4\mu_5} d_L)
                           - 256 Q_2^{qq'} -20 \tilde{E}_2^{qq'(1)} \,, \\
  \tilde{E}_1^{qq'(3)} & = (\bar s_L  \gamma_{\mu_1\mu_2\mu_3\mu_4\mu_5\mu_6\mu_7} T^a q_L)\otimes
                           (\bar q_L' \gamma^{\mu_1\mu_2\mu_3\mu_4\mu_5\mu_6\mu_7} T^a d_L)
                           - 4096 Q_1^{qq'} -336 \tilde{E}_1^{qq'(1)} \,,\\
  \tilde{E}_2^{qq'(3)} & = (\bar s_L  \gamma_{\mu_1\mu_2\mu_3\mu_4\mu_5\mu_6\mu_7} q_L)\otimes
                           (\bar q_L' \gamma^{\mu_1\mu_2\mu_3\mu_4\mu_5\mu_6\mu_7} d_L) 
                           - 4096 Q_2^{qq'} -336 \tilde{E}_2^{qq'(1)} \,.
\end{align*}

We find the following four-loop anomalous dimensions in the
``standard'' basis:
\begin{align}
\begin{split}
\gamma_{Q_1^{qq'} \to Q_1^{qq'}}^{(3)} & = 
       - \frac{11408471}{1296}
       + \frac{23695}{18} N_f
       + \frac{15239}{243} N_f^2
       + \frac{104}{81} N_f^3
\\&
       + \bigg(
            \frac{1116299}{81}
          + \frac{66008}{27} N_f
          - \frac{2240}{27} N_f^2
          - \frac{64}{27} N_f^3
          \bigg) \zeta_3
\\&
       - \bigg(
            \frac{616}{15}
          + \frac{328}{45} N_f
          - \frac{16}{27} N_f^2
          \bigg) \pi^4
       + \bigg(
            \frac{660220}{81}
          - \frac{71920}{27} N_f
          \bigg) \zeta_5\,,
\end{split}
\\[1em]
\begin{split}
\gamma_{Q_1^{qq'} \to Q_2^{qq'}}^{(3)} & =
        \frac{14154631}{1944}
          - \frac{159862}{243} N_f
          + \frac{15272}{729} N_f^2
          - \frac{568}{243} N_f^3
\\&
       - \bigg(
            \frac{190262}{27}
          + \frac{39920}{81} N_f
          + \frac{128}{81} N_f^2
          - \frac{128}{81} N_f^3
          \bigg) \zeta_3
\\&
       + \bigg(
            \frac{2464}{45}
          + \frac{16}{5} N_f
          - \frac{32}{81} N_f^2
          \bigg) \pi^4
       - \bigg(
            \frac{741160}{81}
          - \frac{179680}{81} N_f
          \bigg) \zeta_5\,,
\end{split}
\\[1em]
\begin{split}
\gamma_{Q_2^{qq'} \to Q_1^{qq'}}^{(3)} & =
       - \frac{36723737}{432}
          + \frac{224059}{27} N_f
          - \frac{12632}{81} N_f^2
          - \frac{284}{27} N_f^3
\\&
       + \bigg(
            \frac{173789}{3}
          - \frac{116872}{9} N_f
          + \frac{1664}{9} N_f^2
          + \frac{64}{9} N_f^3
          \bigg) \zeta_3
\\&
       + \bigg(
            \frac{1232}{5}
          + \frac{72}{5} N_f
          - \frac{16}{9} N_f^2
          \bigg) \pi^4
       - \bigg(
            \frac{370580}{9}
          - \frac{89840}{9} N_f
          \bigg) \zeta_5\,,
\end{split}
\\[1em]
\begin{split}
\gamma_{Q_2^{qq'} \to Q_2^{qq'}}^{(3)} & =
       - \frac{4310269}{324}
          + \frac{67852}{81} N_f
          - \frac{1624}{81} N_f^2
\\&
       + \bigg(
            \frac{773336}{81}
          - \frac{4352}{27} N_f
          \bigg) \zeta_3
\\&
       + \bigg(
            \frac{616}{15}
          - \frac{112}{45} N_f
          \bigg) \pi^4
       - \bigg(
            \frac{451520}{81}
          - \frac{17920}{27} N_f
          \bigg) \zeta_5\,.
\end{split}
\end{align}
We have checked these results by direct calculation in the standard
operator basis. We also reproduced the one-, two-, and three-loop
anomalous dimensions of the $Q_1^{qq'}-Q_2^{qq'}$ sector from
Ref.~\cite{Gorbahn:2004my}. This constitutes the first independent
check of the three-loop anomalous dimension (in the current-current
sector) in Ref.~\cite{Gorbahn:2004my}.

\section{Conclusions\label{sec:conclusions}}

We have performed the first calculation of the anomalous dimension of
the $|\Delta S| = 2$ current-current operators at NNNLO in QCD. This
calculation is the first step in the prediction of the charm-top
contribution to $\epsilon_K$ at NNNLO that we will present in
subsequent publications. Our results are given in fully analytic form
and have been obtained in a basis of evanescent operators that ensures
that the anomalous-dimension matrix in the physical sector is diagonal
up to the four-loop level. We have also provided all expressions that
are necessary to transform our results into any other operator basis
that is obtained by adding positive powers of $\epsilon$ times
physical operators to the evanescent operators. We explicitly
displayed the result corresponding to the ``standard'' operator basis
used in $B$ physics~\cite{Gorbahn:2004my}. Moreover, we have
reproduced the three-loop anomalous dimension for the operators
$Q_1^{qq'}$ and $Q_2^{qq'}$, presented in Ref.~\cite{Gorbahn:2004my},
by direct calculation for the first time and find full agreement.

\section*{Acknowledgments}

J.B. thanks his colleagues from the LHCb group at U Cincinnati for
providing computing resources and acknowledges support by DoE grant
DE-SC0011784.
E.S.~and T.S.~acknowledge partial funding by the Deutsche Forschungsgemeinschaft (DFG, German Research Foundation) 
under Germany’s Excellence Strategy – Cluster of Excellence ``Color meets Flavor'', EXC 3107 – Project-ID 533766364.

\appendix

\section{Renormalization constants\label{sec:Z}}

Various field, mass, and coupling renormalization constants up to four-loop level
enter our calculation; we have calculated all of them explicitly. 
We denote them by $Z_r$, with $r =
q,g_s,G,u,m_{\text{IRA}}$ denoting the quark field, strong
coupling, gluon field, ghost field, and artificial gluon mass
renormalization, respectively, and write them as an expansion in
analogy to Eq.~\eqref{eq:def:Z:exp}. The gluon mass counterterm arises
because the use of infrared rearrangement~\cite{Chetyrkin:1997fm}, which we employ, 
breaks gauge invariance in intermediate steps of the calculation. At the
renormalizable level this method generates one gauge-non-invariant operator
corresponding to a gluon-mass term, i.e.,
\begin{equation}
  {\mathcal L} \supset \frac{1}{2} Z_{\text{IRA}} m_{\text{IRA}}^2 G_\mu^a G^{\mu,\,a} \,.
\end{equation}
The effective gluon mass, $m_{\text{IRA}}$, is completely artificial
and drops out of all physical results, and $Z_{\text{IRA}}$ is an
additional renormalization constant~\cite{Chetyrkin:1997fm}. No
further, non-renormalizable counterterms related to the infrared
rearrangement are required in our calculation.

For convenience, we display all our results as a polynomial in
$\xi-1$, and find by explicit calculation
\begin{align}
  Z_q^{(1,1)}           & = - \frac{4}{3} - \frac{4}{3} (\xi-1) \,,\\
  Z_{g_s}^{(1,1)}        & = - \frac{11}{2} + \frac{1}{3} N_f\,, \\
  Z_G^{(1,1)}           & =   5 - \frac{2}{3} N_f
                           - \frac{3}{2} (\xi - 1)\,, \\
  Z_u^{(1,1)}           & = \frac{3}{2} - \frac{3}{4} (\xi-1) \,,\\
  Z_{{\text{IRA}}}^{(1,1)} & = - 3 - 2 N_f
                               - \frac{9}{4} (\xi-1) \,,\\[1em]
  Z_q^{(2,2)}           & =   \frac{44}{9}
                           + \frac{61}{9} (\xi-1)
                           + \frac{17}{9} (\xi-1)^2 \,, \\
  Z_{g_s}^{(2,2)}        & = \frac{363}{8} - \frac{11}{2} N_f + \frac{1}{6} N_f^2\,, \\
  Z_G^{(2,2)}           & =   \frac{5}{2} N_f - \frac{75}{4}
                           + \bigg( N_f - \frac{15}{8} \bigg) (\xi-1)
                           + \frac{9}{4} (\xi-1)^2 \,, \\
  Z_u^{(2,2)}           & =   \frac{3}{4} N_f - 9
                           + \frac{27}{16} (\xi-1)
                           + \frac{27}{32} (\xi-1)^2 \,,\\
  Z_{{\text{IRA}}}^{(2,2)} & =   \frac{81}{4} + \frac{101}{12} N_f
                            + \bigg( \frac{261}{16} + \frac{29}{12} N_f \bigg) (\xi-1)
                            + \frac{135}{32} (\xi-1)^2 \,, \\[1em]
  Z_q^{(2,1)}           & = - \frac{47}{3} + \frac{2}{3} N_f
                           - 5 (\xi-1)
                           - \frac{1}{2} (\xi-1)^2 \,, \\
  Z_{g_s}^{(2,1)}           & = \frac{19}{6} N_f - \frac{51}{2} \,, \\
  Z_G^{(2,1)}           & =  \frac{207}{8} - \frac{61}{12} N_f
                          - \frac{135}{16} (\xi-1)
                          - \frac{9}{8} (\xi-1)^2 \,, \\
  Z_u^{(2,1)}           & =   \frac{147}{16} - \frac{5}{8} N_f
                           + \frac{9}{32} (\xi-1) \,,\\
  Z_{{\text{IRA}}}^{(2,1)} & = - \frac{219}{16} - \frac{59}{24} N_f
                            - \bigg( \frac{29}{24} N_f + \frac{441}{32} \bigg) (\xi-1)
                            - \frac{45}{16} (\xi-1)^2 \,, \\[1em]
\begin{split}
Z_q^{(3,1)}           & =   \frac{1253}{54} N_f - \frac{20}{81} N_f^2 - \frac{24941}{108} + \frac{26}{3} \zeta_3
                         + \bigg( \frac{17}{6} N_f - \frac{371}{8} - 6 \zeta_3 \bigg) (\xi-1) \\
& \quad                  - \bigg( \frac{69}{8} + \frac{3}{2} \zeta_3 \bigg) (\xi-1)^2
                         - \frac{5}{4} (\xi-1)^3 \,,
\end{split}\\
Z_{g_s}^{(3,1)}           & = - \frac{2857}{12} + \frac{5033}{108} N_f
                            - \frac{325}{324} N_f^2 \,, \\
\begin{split}
  Z_G^{(3,1)}           & =   \frac{4051}{16} - \frac{7831}{108} N_f + \frac{215}{81} N_f^2
                           + \bigg( 11 N_f - \frac{27}{2} \bigg) \zeta_3 \\
& \quad                    + \bigg( 3 N_f - \frac{1143}{16} - \frac{81}{8} \zeta_3 \bigg) (\xi-1)
                           - \bigg( \frac{243}{16} + \frac{27}{16} \zeta_3 \bigg) (\xi-1)^2
                           - \frac{63}{32} (\xi-1)^3 \,,
\end{split}\\
\begin{split}
  Z_u^{(3,1)}           & =   \frac{229}{3} - \frac{1085}{144} N_f - \frac{35}{108} N_f^2
                           - \bigg( \frac{11}{2} N_f - \frac{27}{4} \bigg) \zeta_3 \\
& \quad                    + \bigg( \frac{21}{16} N_f - \frac{495}{64} + \frac{81}{16} \zeta_3 \bigg) (\xi-1)
                           - \bigg( \frac{135}{64} - \frac{27}{32} \zeta_3 \bigg) (\xi-1)^2
                           - \frac{27}{64} (\xi-1)^3 \,,
\end{split}\\
\begin{split}
  Z_\text{IRA}^{(3,1)}   & = - \frac{22837}{192} + \frac{317}{108} N_f + \frac{77}{108} N_f^2
                           + \bigg( 5 N_f - \frac{81}{8} \bigg) \zeta_3 \\
& \quad                    - \bigg[ \frac{7533}{64} - \frac{1045}{216} N_f
                                    + \bigg( \frac{297}{16} + \frac{41}{4} N_f \bigg) \zeta_3 \bigg] (\xi-1) \\
& \quad                    - \bigg( \frac{657}{16} + \frac{599}{1728} N_f + \frac{135}{32} \zeta_3 \bigg) (\xi-1)^2
                           - \frac{441}{64} (\xi-1)^3 \,,
\end{split}\\[1em]
\begin{split}
Z_q^{(3,2)}           & =   \frac{8}{27} N_f^2 - \frac{344}{27} N_f + \frac{1348}{9}
                         + \bigg( \frac{779}{9} - \frac{26}{9} N_f \bigg) (\xi-1) \\
& \quad                  + \frac{125}{6} (\xi-1)^2
                         + \frac{13}{6} (\xi-1)^3 \,,
\end{split}\\
Z_{g_s}^{(3,2)}       & =   \frac{2057}{4} - \frac{3421}{36} N_f
                        + \frac{209}{54} N_f^2 \,, \\
\begin{split}
  Z_G^{(3,2)}           & = - \frac{4325}{16} + \frac{4093}{72} N_f - \frac{91}{54} N_f^2 \\
& \quad                    + \bigg( \frac{85}{8} N_f + \frac{27}{32} \bigg) (\xi-1)
                           + \bigg( \frac{423}{16} + \frac{3}{4} N_f \bigg) (\xi-1)^2
                           + \frac{63}{16} (\xi-1)^3 \,,
\end{split}\\
\begin{split}
  Z_u^{(3,2)}           & = - \frac{1873}{16} + \frac{397}{24} N_f - \frac{5}{18} N_f^2 \\
& \quad                    + \bigg( \frac{549}{64} - \frac{3}{32} N_f \bigg) (\xi-1)
                           + \frac{567}{128} (\xi-1)^2
                           + \frac{9}{16} (\xi-1)^3 \,,
\end{split}\\
\begin{split}
  Z_\text{IRA}^{(3,2)}   & =   \frac{1823}{8} + \frac{12407}{432} N_f - \frac{4}{3} N_f^2\\
& \quad                    + \bigg( \frac{837}{4} + \frac{19439}{864} N_f \bigg) (\xi-1)
                           + \bigg( \frac{9855}{128} + \frac{2243}{432} N_f \bigg) (\xi-1)^2
                           + \frac{693}{64} (\xi-1)^3 \,,
\end{split}\\[1em]
Z_q^{(3,3)}           & =   \frac{2}{3} N_f - \frac{2165}{81}
                         + \bigg( \frac{2}{3} N_f - \frac{2197}{54} \bigg) (\xi-1)
                         - \frac{901}{54} (\xi-1)^2
                         - \frac{221}{81} (\xi-1)^3 \,, \\
Z_{g_s}^{(3,3)}           & = - \frac{6655}{16} + \frac{605}{8} N_f - \frac{55}{12} N_f^2
                              + \frac{5}{54} N_f^3 \,, \\
\begin{split}
  Z_G^{(3,3)}           & =   \frac{825}{8} - \frac{65}{4} N_f + \frac{1}{3} N_f^2 \\
& \quad                    + \bigg( \frac{405}{16} - \frac{27}{4} N_f \bigg) (\xi-1)
                           - \bigg( \frac{45}{8} + \frac{3}{2} N_f \bigg) (\xi-1)^2
                           - \frac{27}{8} (\xi-1)^3 \,,
\end{split}\\
  Z_u^{(3,3)}           & =   \frac{1029}{16} - \frac{19}{2} N_f + \frac{1}{3} N_f^2
                           - \bigg( \frac{9}{2} + \frac{3}{16} N_f \bigg) (\xi-1)
                           - \frac{81}{16} (\xi-1)^2
                           - \frac{135}{128} (\xi-1)^3 \,,\\
\begin{split}
  Z_\text{IRA}^{(3,3)}   & = - \frac{2577}{16} - \frac{10739}{216} N_f + \frac{5}{3} N_f^2\\
& \quad                    - \bigg( \frac{4077}{32} + \frac{8585}{432} N_f \bigg) (\xi-1)
                           - \bigg( \frac{2997}{64} + \frac{1795}{432} N_f \bigg) (\xi-1)^2
                           - \frac{945}{128} (\xi-1)^3 \,,
\end{split}\\[1em]
\begin{split}
Z_q^{(4,1)}           & = - \frac{19684159}{5184} + \frac{53713}{96} N_f
                         - \frac{10483}{972} N_f^2 - \frac{35}{243} N_f^3
                         - \bigg( \frac{167}{120} - \frac{3}{20} N_f \bigg) \pi^4 \\
& \quad                  + \bigg( \frac{67469}{648} + \frac{2653}{54} N_f
                                  - \frac{52}{9} N_f^2 \bigg) \zeta_3
                         + \bigg( \frac{32345}{81} + \frac{40}{3} N_f \bigg) \zeta_5 \,,
\end{split}\\
Z_q^{(4,2)}           & =   \frac{8385571}{2592} - \frac{619771}{1296} N_f
                         + \frac{4202}{243} N_f^2 - \frac{10}{81} N_f^3
                         - \bigg( \frac{1387}{18} + \frac{17}{3} N_f \bigg) \zeta_3 \,, \\
Z_q^{(4,3)}           & = - \frac{198337}{144} + \frac{117661}{648} N_f
                         - \frac{683}{81} N_f^2 + \frac{4}{27} N_f^3 \,, \\
Z_q^{(4,4)}           & = \frac{347125}{1944} - \frac{413}{36} N_f + \frac{2}{9} N_f^2 \,,
\end{align}
where $\xi$ is the gauge fixing parameter in generalized $R_\xi$ gauge
for the gluon fields. Our two-loop renormalization constants agree
with the results in the literature~\cite{Gambino:2003zm} if one bears
in mind that the original papers contain some typographical
errors. Due to the increased complexity of the calculation, we have
calculated the four-loop quark field renormalization only for $\xi =
1$.

Finally, we provide the explicit expressions for the $Z$ factors that
are necessary to perform the change of scheme into any different
operator basis via Eq.~\eqref{eq:gentmat}. In the following, the
matrix entries correspond to the ordering of Wilson coefficients in
the physical sector,
\begin{equation}
  Q \sim \big(C_+^{qq'}, C_-^{qq'}\big)
\end{equation}
and the evanescent sector,
\begin{equation}
  E \sim \big(E_1^{qq',(1)}, E_2^{qq',(1)}, E_1^{qq',(2)}, E_2^{qq',(2)},
          E_1^{qq',(3)}, E_2^{qq',(3)}, E_1^{qq',(4)}, E_2^{qq',(4)}\big) \,.
\end{equation}
For the mixing of physical into evanescent operators we find, at
one-loop,
\begin{equation}
Z_{QE}^{(1,1)} =
\begin{pmatrix}
\frac{13}{12}&\frac{2}{9}&0&0&0&0&0&0\\[0.5em]
-\frac{1}{12}&-\frac{2}{9}&0&0&0&0&0&0
\end{pmatrix}\,,
\end{equation}
at two-loop,
\begin{equation}
Z_{QE}^{(2,1)} =
\begin{pmatrix}
\frac{1105}{72}-\frac{13}{216}N_f &
   \frac{197}{54}-\frac{1}{81}N_f &
   -\frac{137}{1152} & -\frac{91}{864} & 0 & 0 & 0 & 0 \\[0.5em]
 \frac{1}{216}N_f-\frac{187}{72} &
   \frac{1}{81}N_f+\frac{2}{27} &
   -\frac{73}{1152} & \frac{7}{864} & 0 & 0 & 0 & 0
\end{pmatrix}\,,
\end{equation}
and at three-loop,
\begin{align}
  Z_{Q_+^{qq'} \to E_1^{qq'(1)}}^{(3,1)}
& = \frac{40214647}{194400} - \frac{2935}{1944} N_f - \frac{169}{972} N_f^2
    + \bigg( \frac{4669}{81} - \frac{130}{27} N_f \bigg) \zeta_3 \,, \\
  Z_{Q_+^{qq'} \to E_2^{qq'(1)}}^{(3,1)}
& = \frac{23757301}{291600} - \frac{185}{1458} N_f - \frac{26}{729} N_f^2
    - \bigg( \frac{2128}{243} + \frac{80}{81} N_f \bigg) \zeta_3 \,, \\
  Z_{Q_+^{qq'} \to E_1^{qq'(2)}}^{(3,1)}
& = - \frac{91603}{15552} - \frac{1205}{15552} N_f - \frac{427}{162} \zeta_3 \,, \\
  Z_{Q_+^{qq'} \to E_2^{qq'(2)}}^{(3,1)}
& = - \frac{40169}{46656} - \frac{247}{11664} N_f - \frac{161}{486} \zeta_3 \,, \\
  Z_{Q_+^{qq'} \to E_1^{qq'(3)}}^{(3,1)}
& = \frac{24209}{124416} - \frac{25}{324} \zeta_3 \,, \\
  Z_{Q_+^{qq'} \to E_2^{qq'(3)}}^{(3,1)}
& = \frac{179}{186624} + \frac{97}{3888} \zeta_3 \,, \\
  Z_{Q_-^{qq'} \to E_2^{qq'(1)}}^{(3,1)}
& = - \frac{7975187}{97200} - \frac{3227}{1944} N_f + \frac{13}{972} N_f^2
    - \bigg( \frac{1138}{81} - \frac{10}{27} N_f \bigg) \zeta_3 \,, \\
  Z_{Q_-^{qq'} \to E_2^{qq'(1)}}^{(3,1)}
& = - \frac{7141321}{291600} + \frac{71}{1458} N_f + \frac{26}{729} N_f^2
    + \bigg( \frac{5368}{243} + \frac{80}{81} N_f \bigg) \zeta_3 \,, \\
  Z_{Q_-^{qq'} \to E_1^{qq'(2)}}^{(3,1)}
& = \frac{22631}{31104} + \frac{635}{15552} N_f + \frac{299}{324} \zeta_3 \,, \\
  Z_{Q_-^{qq'} \to E_2^{qq'(2)}}^{(3,1)}
& = - \frac{1519}{46656} + \frac{19}{11664} N_f - \frac{289}{486} \zeta_3 \,, \\
  Z_{Q_-^{qq'} \to E_1^{qq'(3)}}^{(3,1)}
& = \frac{2741}{62208} - \frac{115}{2592} \zeta_3 \,, \\
  Z_{Q_-^{qq'} \to E_2^{qq'(3)}}^{(3,1)}
  & = \frac{4441}{186624} - \frac{37}{3888} \zeta_3 \,.
\end{align}
The mixing among the evanescent operators is given at one-loop by
\begin{equation}
Z_{EE}^{(1,1)} =
\begin{pmatrix}
-\frac{46}{3} & -\frac{52}{9} & \frac{5}{12} &
   \frac{2}{9} & 0 & 0 & 0 & 0 \\[0.5em]
 -26 & 0 & 1 & 0 & 0 & 0 & 0 & 0 \\[0.5em]
 -\frac{1600}{3} & -128 & \frac{38}{3} &
   \frac{4}{3} & \frac{5}{12} & \frac{2}{9} & 0 & 0 \\[0.5em]
 -576 & \frac{1280}{3} & 6 & -\frac{64}{3} & 1 & 0 & 0 & 0 \\[0.5em]
 -11264 & -\frac{22528}{9} & -\frac{1792}{3} &
   -\frac{896}{9} & \frac{166}{3} & \frac{76}{9} &
   \frac{5}{12} & \frac{2}{9} \\[0.5em]
 -11264 & \frac{28672}{3} & -448 & \frac{1792}{3} &
   38 & -64 & 1 & 0 \\[0.5em]
 * & * & * & * & * & * & * & * \\[0.5em]
 * & * & * & * & * & * & * & *
\end{pmatrix}\,,
\end{equation}
and at two-loop by
\begin{equation}
\begin{split}
Z_{EE}^{(2,1)} & =
\left(
\begin{matrix}
\frac{143}{27} N_f-\frac{937}{36} &
   \frac{134}{81} N_f-\frac{3619}{54} &
   \frac{257}{72}-\frac{5}{216} N_f &
   \frac{44}{27}-\frac{1}{81} N_f \\[0.5em]
 \frac{67}{9} N_f-\frac{1963}{12} &
   \frac{787}{9}-\frac{32}{9} N_f &
   \frac{89}{6}-\frac{1}{18} N_f &
   \frac{43}{18} \\[0.5em]
 \frac{3560}{27} N_f+\frac{45236}{9} &
   \frac{320}{9} N_f-\frac{5200}{27} &
   \frac{182}{27} N_f+\frac{10393}{36} &
   \frac{14}{9} N_f-\frac{7703}{54} \\[0.5em]
 160 N_f-\frac{25496}{3} &
   \frac{29632}{3}-\frac{2560}{27} N_f & 7
   N_f+\frac{6985}{12} &
   \frac{1667}{9}-\frac{160}{27} N_f \\
 * & * & * & * \\
 * & * & * & * \\
 * & * & * & * \\
 * & * & * & *
\end{matrix}
\right.\\
&\hspace{16em}\left.
\begin{matrix}
   \frac{1}{384} & -\frac{35}{864} & 0 & 0 \\[0.5em]
   -\frac{35}{192} & -\frac{7}{72}
   & 0 & 0 \\[0.5em]
   \frac{131}{72}-\frac{5}{216} N_f &
   \frac{41}{18}-\frac{1}{81} N_f &
   \frac{1}{384} & -\frac{35}{864} \\[0.5em]
   \frac{71}{4}-\frac{1}{18} N_f &
   -\frac{61}{18} & -\frac{35}{192} & -\frac{7}{72} \\
 * & * & * & * \\
 * & * & * & * \\
 * & * & * & * \\
 * & * & * & *
\end{matrix}
\right)\,.
\end{split}
\end{equation}
The asterisks correspond to entries that are not needed for the scheme
transformation up to the four-loop level.

\addcontentsline{toc}{section}{References}
\bibliographystyle{JHEP}
\bibliography{references}

\end{document}